\documentclass[apsrev4-1,twocolumn,prl,showpacs,floatfix,superscriptaddress,citeautoscript]{revtex4-1}
\usepackage{epsf,color,colordvi}
\usepackage{graphicx}
\usepackage{amsmath}
\usepackage{times}
\usepackage{subfigure}
\usepackage{ulem}  
\usepackage{placeins}

\begin{document}
\title{The universal equation of state of a unitary fermionic gas} 

\author{R.K. Bhaduri}
\affiliation{Department of Physics and Astronomy, 
McMaster University, Hamilton, Ontario \L8S 4M1, Canada}
\author{W. van Dijk}
\affiliation{Department of Physics and Astronomy, 
McMaster University, Hamilton, Ontario \L8S 4M1, Canada}
\affiliation{Physics Department, Redeemer University College, 
Ancaster, Ontario \ L9K 1J4, Canada}
\author{M. V. N. Murthy}
\affiliation{The Institute of Mathematical Sciences, Chennai 600113, 
India}
\begin{abstract}
It is suggested that for a fermi gas at unitarity, the two-body bond plays 
a special role. We propose an equation of state using an ansatz relating 
the interaction part of the $l$-body cluster to its two-body counterpart. 
This allows a parameter-free comparison with the recently measured 
equation of state by the ENS group. The agreement between the two over a 
range of fugacity ($z<5$ for a homogeneous gas, and $z<10$ for the trapped 
gas) leads us to perform the calculations of more sensitive quantities 
measured recently by the MIT group.
\end{abstract}


\date{\today} 

\maketitle

Feshbach resonance makes it possible to adjust the strength of the 
inter-atomic interaction in a neutral atomic gas. When the scattering 
length goes to $\pm \infty$, there is no length scale left other than 
the average inter-particle distance and the thermal wavelength (assuming a zero-range interaction). 
The gas is then termed ``unitary'' and its properties are universal 
when expressed in appropriate dimensionless units at all scales whether 
the system is fermionic or bosonic~\cite{ho}. Recent accurate 
measurements by the ENS group~\cite{1,2} and the Tokyo group~\cite{hori} 
have confirmed the universal nature of the equation of state (EOS) of 
a gas of neutral fermionic atoms, and have given fresh impetus to its 
theoretical understanding~\cite{3,3p}. More recently, direct measurements 
by the MIT group~\cite{ku} of the isothermal compressibility $\kappa$, 
pressure $P$, and heat capacity $C_V/N k_B$ for a unitary gas have 
revealed the superfluid transition at $T_c/T_F=0.167 (13)$.

In this paper, keeping in mind the fundamental nature of the two-particle 
bond at unitarity, we propose a description of the unitary gas as 
consisting of singlet pairs, in terms of which all higher order clusters 
are expressed. The resulting Equation of State (EOS) extends the agreement 
with the ENS data~\cite{1,2} on the grand potential over a much larger 
range of fugacity $z$ than expected. However, this description breaks down 
for $z>5$ for the homogeneous gas (and $z>10$ for the harmonically trapped 
gas). For the homogeneous gas, $z=5$ corresponds to a temperature 
$T/T_F=0.22$, below which the proposed EOS cannot be trusted. We 
calculate, with our higher virial coefficients, the pressure, 
compressibility and heat capacity of the homogeneous gas to compare with 
the MIT data~\cite{3}. The calculation of these quantities is a stringent 
test since they require higher moments of the virial expansion.  We find 
that inclusion of the higher virial coefficients yields agreement with the 
MIT data for pressure and entropy down to $T/T_F =0.3$, and the 
compressibility and heat capacity to $T/T_F=0.6$.

To set the stage for the proposed universal EOS, we briefly recapitulate 
the virial expansion of a two-component interacting homogeneous fermi 
gas~\cite{4}.  The grand potential $\Omega(\beta,\mu)$ is defined as 
$\Omega=-\tau \ln {\cal Z}$, where $\tau=k_BT=1/\beta$ and ${\cal Z}$ is 
the grand-canonical partition function.  Furthermore $\Omega = -PV$, and 
may be expressed in a power series of fugacity $z=\exp(\beta\mu)$, where  
$\mu$ is chemical potential. The grand potential $\Omega = -\tau 
Z_1(\beta)\sum_{l=1}^\infty 
b_lz^l$, 
where $Z_1(\beta)$ is the one-body partition function, and $b_l$ is
the $l$-particle cluster integral.
For an untrapped gas in volume $V$, we have $Z_1(\beta)=2
(V/\lambda^3)$, where spin degeneracy of $2$ is included and 
$\lambda=(2\pi\hbar^2\beta/m)^{1/2}$ is the thermal wave length.  
For a harmonic oscillator (HO) trap in three dimensions, 
$Z_1(\beta)=2/(\hbar\omega\beta)^3$.
For a unitary gas, the cluster integrals $b_l$'s are also temperature 
independent in the high-temperature expansion.  Subtracting from $\Omega$ 
the ideal part of the grand potential $\Omega^{(0)}$ we obtain the 
interaction part of the EOS as 
\begin{equation}\label{eq:3}
\Omega -\Omega^{(0)} = -\tau Z_1(\beta)\sum_{l=2}^\infty (\Delta b_l)z^l,
\end{equation}
where $\Delta b_l = b_l - b_l^{(0)}.$  Note that $b_1=b_1^{(0)}=1$, and 
cancels out on taking the difference.  
 
Consider now the special role played by $\Delta b_2$ of the two-particle 
cluster at unitarity.  In such a gas, the spin-up fermions have a tendency 
to pair up with the spin-down fermions because the short-range interaction 
potential is on the verge of producing zero-energy bound states.  
The Feshbach resonance being in the relative $s$-state, ensures the 
pair interaction to be operative only between singlet pairs.  
One finds that~\cite{5} 
$\Delta b_2 = (2\sqrt{2})\times {1\over 2}(\Delta Z_2)$, 
where the factor $2\sqrt{2}$ arises from the CM motion, $\Delta Z_2$ 
is the relative two-body partition function, and the ``suppression factor" 
${1\over 2}$ arises from the fact that only half of the $N$ particles 
can interact in a spin-balanced two-component Fermi gas.  Note that~\cite{6} 
at unitarity $\Delta Z_2 = {1\over 2}$, yielding 
$\Delta b_2 = {1\over\sqrt{2}}$ for such a system.  
What about the $\Delta b_l$'s for the $l$-body clusters that appear 
in Eq.~(\ref{eq:3})?  Keeping in mind that the unitary gas may be looked 
upon as a system consisting of forming and dissolving two-body pairs, 
{\it we conjecture} that for the scale invariant system, the $\Delta 
b_l$ for $l>2$ should be expressible in terms of $(\Delta b_2)$ with 
an appropriate suppression factor.  
Viewing a $l$-body cluster as one particle interacting with the 
rest from a cluster of $(l-1)$ paired particles, we assume that the 
suppression factor is given by $2^{{\cal N}_{(l-1)}}$, where 
${\cal N}_{(l-1)}=(l-1)(l-2)/2$ 
is in general the number of pairs in a cluster with $(l-1)$ fermions.  
Thus our basic ansatz is 
\begin{equation}\label{eq:4}
\Delta b_l = (-)^l \dfrac{(\Delta b_2)}{2^{{\cal N}_{(l-1)}}}, 
\ \ l\geq 2.
\end{equation}
For $l=2, \ {\cal
  N}_1=0$, and Eq.~(\ref{eq:4}) is an identity.  
The alternating sign $(-)^l$ in the above equation was put in to keep
the number fluctuation $(\Delta N)^2/\bar{N} =\sum_l l^2b_l z^l/\sum_ll 
b_lz^l$ not to grow to a very large value with $z$, where $(\Delta
N)^2=\overline{N^2}-{\overline{N}}^2$ 
is the number fluctuation, proportional to the isothermal 
compressibility~\cite{landau}. A large value of the compressibility would lead to a vanishing monopole excitation which is a signature of instability~\cite{bhaduri84}. 

Although our description of the higher virial coefficients in terms 
of the second may seem to be very different from the conventional one, 
similar relationship between the third and second virial coefficients have 
been found in anyons which is also a scale invariant system \cite{7,any}. 
This is obtained by demanding that the divergences in the three-body 
clusters cancel by similar divergences in two-body clusters in the high 
temperature limit. A formal derivation for arbitrary $l$ for the unitary 
gas appears to be non-trivial.

With this ansatz,
\begin{equation}\label{eq:5}
\Omega - \Omega^{(0)} = -\tau Z_1(\beta) (\Delta b_2)\sum_{l=2}^\infty 
(-)^l \dfrac{z^l}{2^{{\cal N}_{(l-1)}}}.
\end{equation}
Experimentally~\cite{1,3,8}, it is the quantity $h(\zeta) 
= \Omega/\Omega^{(0)}$ that is extracted, where $\zeta=1/z$.  
This is given by
\begin{equation}\label{eq:6}
h(\zeta) = 1 + (\Delta b_2)\frac{\sum_{l=2}^\infty (-)^l  
(\zeta)^{-l}/2^{{\cal N}_{(l-1)}}}{\tilde{\Omega}^{(0)}}.
\end{equation}
In a homogeneous gas with a spin-degeneracy of 2,
$$\tilde{\Omega}^{(0)} = \dfrac{2}{\sqrt{\pi}}\int_0^\infty \sqrt{t} 
\ln(1+ze^{-t})\: dt $$
It is worth noting that using Eq.~(\ref{eq:4}) with $\Delta b_2 = 
1/\sqrt{2}$,  we obtain
$\Delta b_3 = -1/2\sqrt{2}, \ \Delta b_4 =1/8\sqrt{2},
\ \Delta b_5 = -1/64\sqrt{2}, \mathrm{etc.}$
Numerically $\Delta b_3$ is known to great accuracy, it was calculated 
up to 8 decimal figures in ~\cite{9} and has 
now been improved to 12 decimal figures ~\cite{rak}. 
Our ansatz for the third virial coefficient differs from the numerically 
computed value in the third decimal and as such cannot be exact. 
However, as we shall see the agreement with EOS data is unaffected by 
such fine differences in $\Delta b_3$.
It is also estimated~\cite{1}  
that $\Delta b_4 \approx 0.096\pm0.015$, and is
consistent with our prediction within the error bars.
It should be mentioned that $\Delta b_4$ as quoted in \cite{rak} is
different sign and magnitude from~\cite{1} and our value.
This however destroys the agreement with the data from the ENS group.

Before confronting the experimental data, we note that for a gas
trapped in a three-dimensional harmonic oscillator (HO), Eqs.~(\ref{eq:6})
is modified to~\cite{1,3}
\begin{equation}\label{eq:9}
h(\zeta) = 1 + (\Delta b_2)\sum_{l=2}^\infty \dfrac{(-)^l}{(l)^{3/2}} 
\dfrac{\zeta^{-l}}{2^{{\cal N}_{(l-1)}}} / \widetilde{\Omega}^{(0)},
\end{equation}
and
$\widetilde{\Omega}^{(0)} = {1\over 2}\int_0^\infty t^2 
\ln\left(1+ze^{-t}\right)\: dt$.
The additional suppression factor of $1/l^{3/2}$ in
Eq.~(\ref{eq:9}) was derived in \cite{9} assuming a local fugacity in 
a HO potential.

\begin{figure}[htbp]
\centering
\resizebox{3.5in}{!}{\includegraphics[angle=-90]{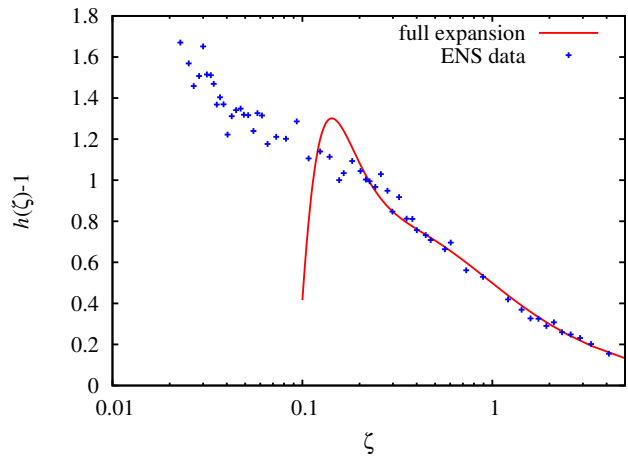}}
\caption{(Colour online) The function $h(\zeta)-1$ for the untrapped unitary Fermi gas (from Eq.~(\ref{eq:6})) as a function of $\zeta$.  The crosses represent the experimental data presented by Nascimb\`ene \textit{et al.}~\cite{1} in their Fig.~3a.}
\label{fig:01}
\end{figure}

\begin{figure}[htbp]
\centering
\resizebox{3.5in}{!}{\includegraphics[angle=-90]{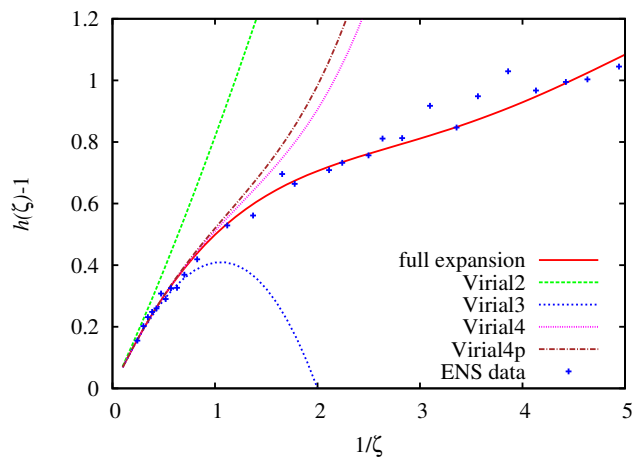}}
\caption{(Colour online) The universal function $h(\zeta)-1$ for the the untrapped fermi gas plotted as a function of the fugacity $1/\zeta$.  The 2nd, 3rd, and 4th virial expansions are also shown.  The 4th order expansion labelled Virial4p has $\Delta b_4 = 0.096$.  The experimental data are the same as in Fig.~\ref{fig:01}.}
\label{fig:02}
\end{figure}

We are now ready to compare our predictions given by Eq.~(\ref{eq:6}) 
for the homogeneous gas and Eq.~(\ref{eq:9}) for the HO with 
experimental data.  In Fig.~\ref{fig:01},
$h(\zeta)-1$ for the homogeneous unpolarized gas, as given by our
Eq.~(\ref{eq:6}), is plotted against $\zeta$.  The crosses on the plot
are the ENS experimental data as found by Nascimb\'ene \textit{et
  al.}~\cite{1} The authors quote that $h$ and $\zeta$ are accurate to
within 6 percent. 
It will be seen from this figure that the series given by our Eq.~(\ref{eq:6}) is in good agreement with the data down to $\zeta\approx 0.2$.  To put the agreement in perspective the same data are plotted as a function of $z=1/\zeta$ in Fig.~\ref{fig:02}, 
along with the behavior of the EOS including virial coefficients up to
the fourth order, as was done by Hu \textit{et al.}~\cite{3}  We see
that such a truncated series could match the data to about $z\approx
1.7$.  Our series~(\ref{eq:6}) extends this to about $z\approx 5$.
This also underlines the importance of higher-order virial
coefficients $\Delta b_l$'s for $l>4$, despite their rapidly
diminishing values.  For the curve labelled Virial4p we set $\Delta
b_4=0.096$, the estimated value~\cite{1}, rather than $0.088$ given by
our ansatz. Note that the $h(\zeta)$ has been calculated in \cite{3p} 
within Pade approximation and including up 
to $\Delta b_3$. Despite deviation from ENS data for $\zeta <1$, they 
obtain surprisingly good agreement for energy and entropy per particle 
down to $T/T_F=0.16$.

\begin{figure}[htbp]
\centering
\resizebox{3.5in}{!}{\includegraphics[angle=-90]{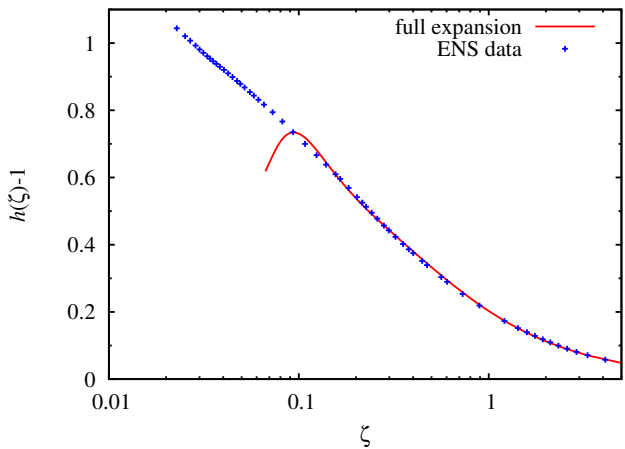}}
\caption{(Colour online) The universal function $h(\zeta)-1$ for 
fermions in an harmonic trap as a function of $\zeta$, Eq.~(\ref{eq:9}).  
The crosses represented the experimental data presented in Fig.~6 of 
Hu \textit{et al.}~\cite{3}}
\label{fig:03}
\end{figure}

\begin{figure}[htbp]
\centering
\resizebox{3.5in}{!}{\includegraphics[angle=-90]{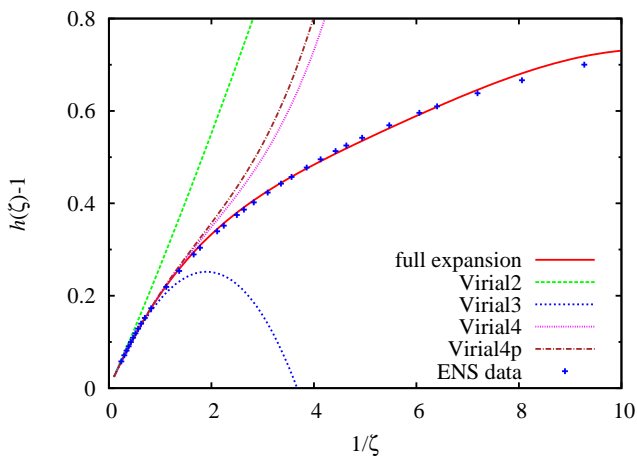}}
\caption{(Colour online) The universal function of trapped fermions 
$h(\zeta)-1$ as a function of fugacity.  See Eq.~(\ref{eq:9}). 
Virial4p is the 4th-order virial expansion with $\Delta b_4 = 0.096$. 
The experimental data are the same as in Fig.~\ref{fig:03}.}
\label{fig:04}
\end{figure}

We now turn to the  ENS measurement for the trapped unitary 
unpolarized gas, as extracted by Hu \textit{et al.}~\cite{3}, and 
compare with our Eq.~(\ref{eq:9}).  (See Fig.~\ref{fig:03}.)
Here the convergence of the virial series is faster as expected, 
and the agreement is remarkably good down to $\zeta\approx 0.1$.   
Figure~\ref{fig:04} 
shows this clearly when the truncated predictions from previous work 
are compared with our result.  The range of applicability of the 
virial series (\ref{eq:9}) is now extended fourfold to $z\approx 10$.  

It should be noted that the series given in Eqs.~(\ref{eq:6}) and 
(\ref{eq:9}) converge for any value of $z$. However, the range of validity 
of the sum depends on the maximum value of $l$ as seen from 
Fig.~\ref{fig:02} and Fig.~\ref{fig:04} for free gas and HO 
respectively. The first few terms make a significant difference but the 
importance of $\Delta b_l$ for $l>8$ is minimal even at $z=5$. The series 
gets saturated by the first twenty terms which is denoted as ``full 
expansion" in all the figures.

Encouraged by the agreement with $h(z)$ even deep into region $z>1$ 
where the normal virial expansion is not expected to work, we next compare 
our 
predictions for the recently measured data on compressibility, heat 
capacity, and pressure by the MIT group~\cite{ku}. Following their 
notation we write $\beta P=f_P(X)/\lambda^3$, where $P$ is the pressure, 
$X=\ln z =\beta\mu$ and $f_P(X)$ is the universal function given by
\begin{equation} 
f_P(X)=2\left(f_{5/2}(\exp(X)) +\sum_{l=2}^{\infty} \Delta b_l\exp(l~ X) 
\right). 
\label{fp}
\end{equation}
The first term in the bracket is the
contribution due to the ideal fermi gas, and is the standard fermi-dirac
integral~\cite{pathria}. All thermodynamic quantities can now be expressed 
in terms of this universal function and its derivatives. Specifically we 
have pressure and compressibility normalized by their zero temperature
values, $P_0,\kappa_0$,  given by
\begin{equation}
\tilde{p}=\frac{P}{P_0}=\frac{5T}{2T_F}\frac{f_P(X)}{f'_P(X)};
~~~\tilde{\kappa}=\frac{\kappa}{\kappa_0}
=\frac{2T_F}{3T}\frac{f''_P(X)}{f'_P(X)},
\label{compress}
\end{equation}
where $T/T_F=4\pi/[3\pi^2f'_P(X)]^{2/3}$ is the dimensionless temperature 
scale and the prime denotes a derivative with respect to $X$. The heat 
capacity at constant volume and entropy are given by
$C_V/Nk_B=\frac{15}{4}\frac{f_P(X)}{f'_P(X)}
-\frac{9}{4}\frac{f'_P(X)}{f''_P(X)}
=\frac{3T_F}{2T}(\tilde{p}-1/\tilde{\kappa}),
S/Nk_B =\frac{5}{2}\frac{f_P(X)}{f'_P(X)}
-\ln(z)$.
The above expressions allow one to calculate the relevant quantities 
either as a function of fugacity $z$ or temperature $T/T_F$ using the
virial expansion given by Eq.~(\ref{fp}) and compare with the respective
experimental data of Ku {\it et al}.

In the light of our earlier remarks (see Fig. 1), these comparisons are 
limited to $z$ values less than $4.95$, corresponding to $T/T_F>0.22$. 
Fig. 5 shows the variation of the pressure, entropy, and heat 
capacity as a function of $T/T_F$. The agreement with experimental data 
improves noticeably as the higher $\Delta b_l$'s are included. 
The agreement for the pressure and entropy hold to 
$T/T_T=0.3$, indicating that the first moments of the virial expansion 
are good.This is not the case for the second moments, however, as the 
plots for heat capacity vs $T/T_F$ shows.  The 
theoretical plots start deviating appreciably from the data for 
$T/T_F<0.6$.
\begin{figure}[htbp]
\centering
\resizebox{3.5in}{!}{\includegraphics[angle=0]{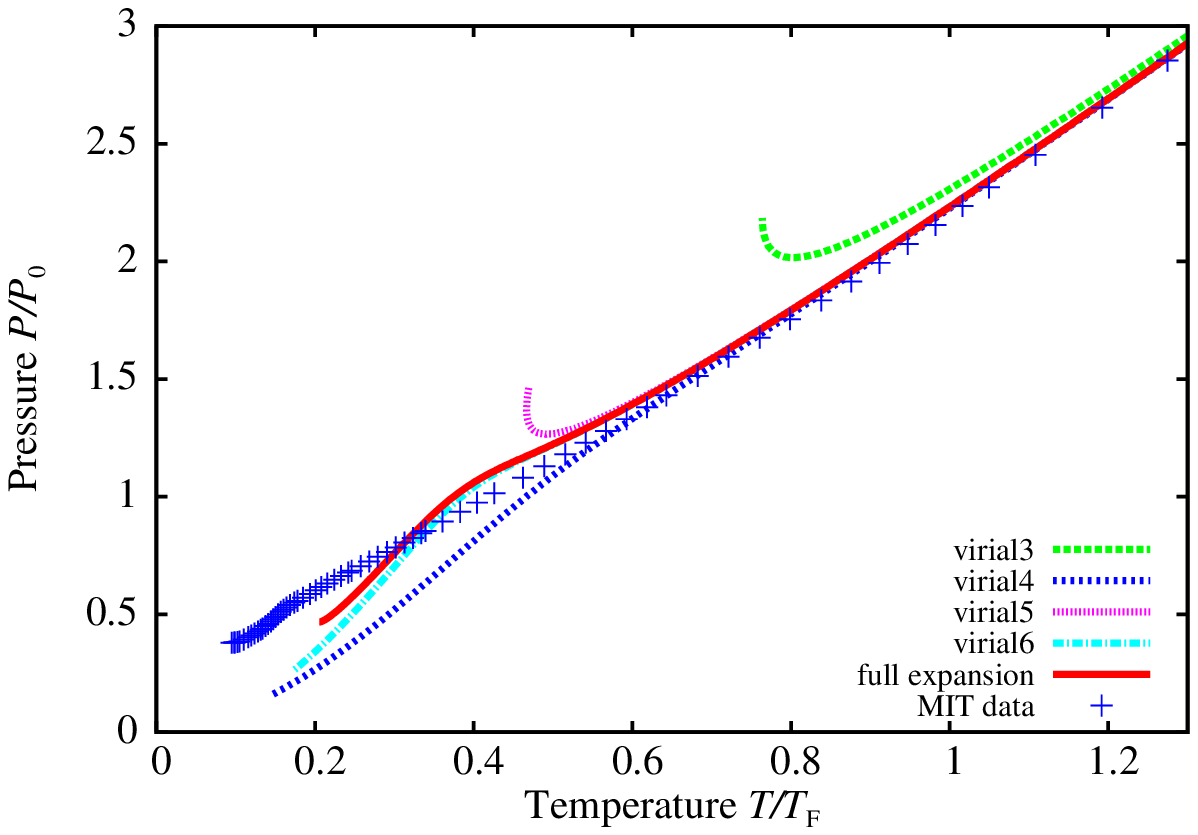}}
\resizebox{3.5in}{!}{\includegraphics[angle=0]{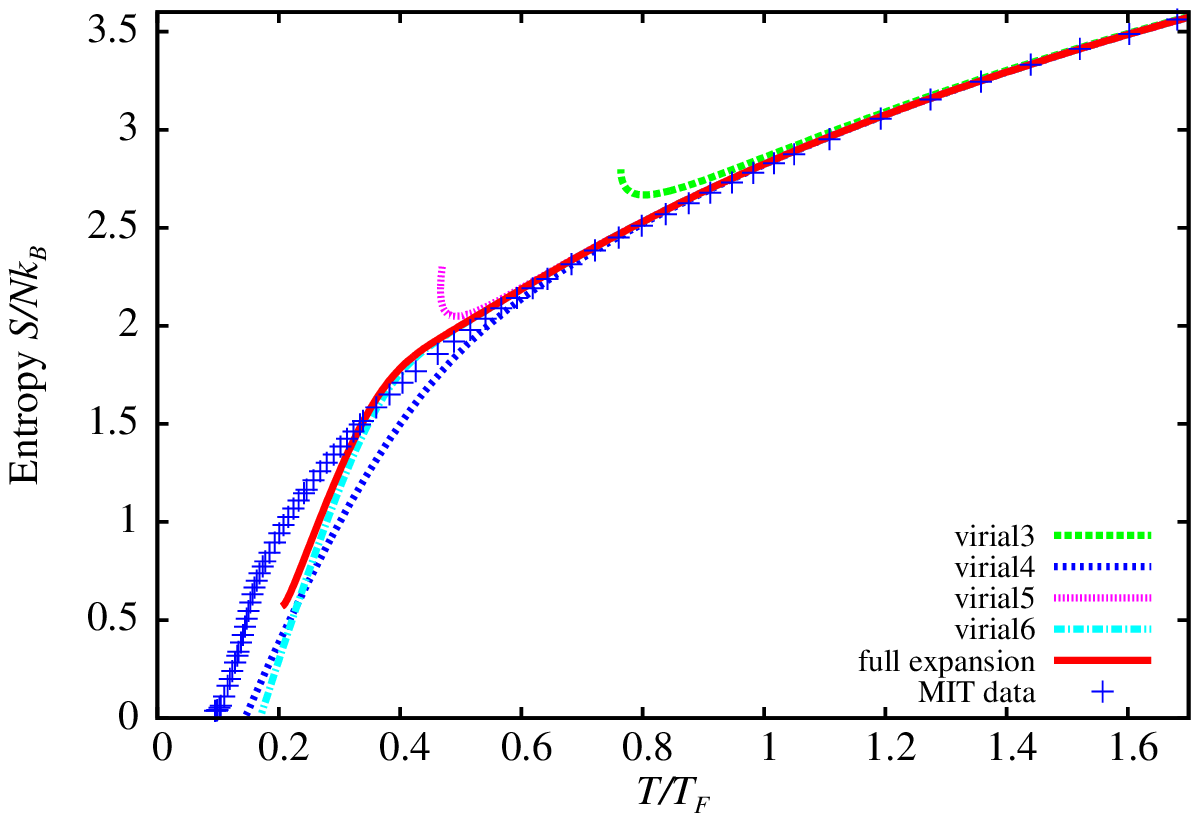}}
\resizebox{3.5in}{!}{\includegraphics[angle=0]{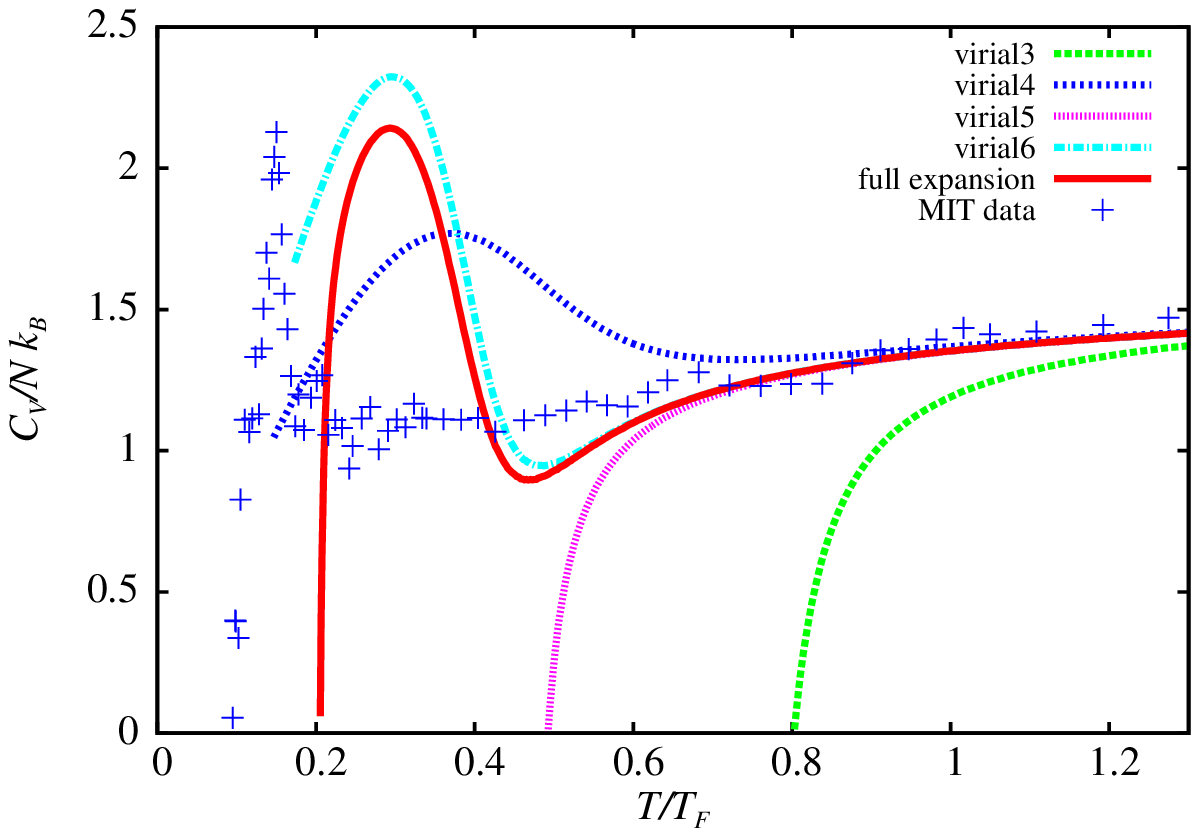}}
\caption{(Colour online) Reduced pressure (top), entropy 
(middle) and heat capacity (bottom) shown as a function of $T/T_F$ for the untrapped unitary Fermi gas.
The experimental data are taken from Ku \textit{et al.}~\cite{ku}.}
\label{fig:05}
\end{figure}
The same behavior is seen in Fig. 6 where the compressibility is plotted 
as a function of pressure (in reduced variables). 
\begin{figure}[htbp]
\centering
\resizebox{3.5in}{!}{\includegraphics[angle=0]{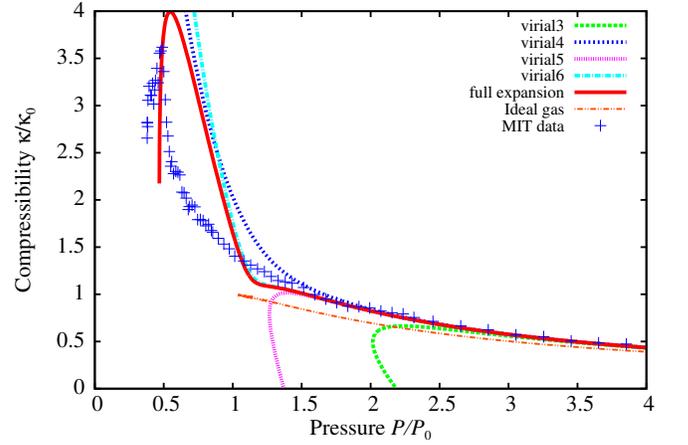}}
\caption{(Colour online) Reduced compressibility  
shown as a function of reduced pressure for the untrapped unitary Fermi gas.
The experimental data are taken from Ku \textit{et al.}~\cite{ku}.}
\label{fig:06}
\end{figure}
It is interesting to note that despite these deviations, a peak in the 
compressibility of about the right magnitude appears in the theoretical 
curve, though at a higher value of $P/P_0$ or $T/T_F$. Though tempting, we 
are reluctant to interpret this as indicative of the onset of 
superfluidity in view of the inaccuracy of the virial description in this 
range of temperature or pressure.

We conclude that the high-temperature virial expansion, in conjunction 
with our ansatz, can match the EOS over a significantly larger range of 
fugacity, corresponding to about $T/T_F\approx 0.3$ for the homogeneous 
gas. Our ansatz (given by Eq.~(\ref{eq:4})) resulted from 
the picture of a unitary fermi gas as a dynamic collection of singlet 
pairs, and assumed that $(\Delta b_2)$ determines the higher virial 
coefficients.  The resulting success of this picture may point to some 
truth in this conjecture, and poses a challenge for deeper understanding.

We thank S. Nascimb\`ene, H. Hu and M. Ku for sharing the experimental 
data for the ideal and trapped fermion gas. We thank S. Das Gupta for helpful discussions.  WvD acknowledges financial support from the Natural 
Sciences and Engineering Research Council of Canada, and MVN acknowledges the 
hospitality of Department of Physics and Astronomy, McMaster University 
where part of this work was done.

\end{document}